      [1999/12/01 v1.4c Il Nuovo Cimento]
\documentclass{cimento}

\usepackage{graphicx}
\title{Recent results and perspectives on cosmic rays ground experiments}
\author{O.~Pisanti\from{ins:x}\ETC}

\instlist{\inst{ins:x} Dipartimento di Scienze Fisiche, Universit\`a di
Napoli ``Federico II'', \atque INFN - Sezione di Napoli - Complesso
Universitario di Monte S.Angelo, Via Cinthia, 80126, Napoli, Italy.
E-mail: \texttt{pisanti@na.infn.it}}

\PACSes{\PACSit{95.55.Ka}{X- and $\gamma$-ray telescopes and
instrumentation} \PACSit{95.55.Vj}{Cosmic ray detectors}
\PACSit{95.85.Pw}{$\gamma$-ray} \PACSit{95.85.Ry}{Cosmic rays}}
\begin{document}

\maketitle

\begin{abstract}
I summarize in this paper the results and perspectives of representative
ground experiments for the observation of very high energy cosmic rays.
\end{abstract}

\section{Introduction}

Although Cosmic Rays (CR's) were first discovered at the beginning of 20th
century, they are still today a subject of intense study since we miss a
complete understanding of their origin, composition and acceleration
mechanisms. Their energies extend from MeV values (with a flux of more
than 1000 particles per second per square meter) to more than $10^{11}$
GeV (less than one particle per km$^2$ per century). In particular, unlike
charged cosmic rays which are deflected by electromagnetic fields, photons
and neutrinos travel almost unimpeded from their sources to Earth;
therefore, their observation is very promising for unveil the details on
production and acceleration mechanisms of CR's. In this paper, we will
focus on Very High Energy (VHE) CR's in the energy range 100 GeV $\div$
100 EeV (1 EeV = $10^{18}$ eV); in this energy region, due to the
smallness of the flux, small detectors on balloons or in outer space are
not adequate, and the atmosphere is usually employed as a big calorimeter
to detect primary particles through the secondaries (charged particles,
fluorescence and Cherenkov light) that they produce hitting on air
molecules. Since, on one side, the dynamic energy range of a detector is
limited to 2-3 orders of magnitude, and we are still considering, on the
other side, an energy range of more than 10 orders of magnitude, we can
expect that different techniques are used at different energies; due to
space limits, we will give details only on some representative experiment,
both at low and high energy.

\section{Low energy experiments}

In the low energy range ($E<10^3$ TeV), two different type of detectors
are used both in $\gamma$-ray astronomy and/or charged CR's detection:
Imaging Atmospheric Cherenkov Telescopes (IACT) or Extensive Air Shower
(EAS) arrays. Both of these techniques take advantage of the fact that,
when a primary CR interacts with the atmosphere, it produces a shower of
relativistic secondary particles which can be all collected or only
sampled by an array of ground detectors in an EAS configuration: the
shower relativistic particles emit also Cherenkov light, that can be
focused by a mirror to a photomultiplier ``camera" of a IACT detector.

In table 1 of ref. \cite{IACT} a list of IACT experiments is given. An
advantage of this approach over the satellite-based one is the collection
area (typically $10^5$ m$^2$, increasing with energy), almost five orders
of magnitude larger than in the case of space detectors. On the other
side, IACT technique differs from the other ground-based approaches in the
better angular resolution, typically $\sim 0.1^\circ$, which is the best
of any astronomical technique above 0.1 MeV, and the lower energy
threshold, $\sim$ 20-50 GeV, almost bridging the gap with space-borne
$\gamma$ experiments, which are limited to $E<$ few GeV. IACT instruments
have also a superior instantaneous sensitivity, defined as the minimum
flux (in \% of that of the Crab Nebula) for which a source is detectable
at a 5$\sigma$ significance in 50 hours of observation. The advantages of
stereoscopic measurements was firstly showed by the HEGRA collaboration
\cite{hegra}: these allow a better reconstruction of direction and energy
and the shower core location can be better established with improved
resolution. A possible limit of IACT detectors is the restricted Field of
View (FoV), $\sim 3-4^\circ$: a consequence of this is, for example, that
in a Gamma Ray Burst (GRB) observation these detectors can only be
operated in ``follow up" mode, requiring a time of the order of minutes
for pointing the object (as an example, the MAGIC \cite{magic1} telescopes
have a slewing speed of $\sim 5^\circ$ s$^{-1}$, about 3 times faster than
HESS's \cite{hess} ones). Moreover, a $\sim 10\%$ duty cycle is due to the
need of astronomical darkness.

With a large FoV ($\sim$2 sr) and a duty cycle of almost 100\%, EAS arrays
are a complementary technique to IACT's for the detection of CR's. Unlike
the experiments of the first generation (CIGNUS \cite{cignus}, CASA
\cite{casa}), where the instrumented area was $\sim$ 1\% of the total one,
implying high energy thresholds, the new generation of experiments (see
table 1 of ref. \cite{lowEAS} for a list of them) used two different
approaches to lower the energy threshold: to instrument a larger fraction
of the total area, in such a way to raise the number of particles arriving
to earth which are detected, and/or locate the detectors at higher
altitude, where it is possible to detect the shower maxima of low energy
showers.

The observations coming from IACT and EAS experiments allow us to shed
light on phenomena taking place in several galactic and extragalactic
sources of CR: pulsars, binary systems, supernovae remnants, Active
Galactic Nuclei (AGN), GRB. In the following, we will report some example
for the two experiments MAGIC \cite{magic1} and ARGO-YBJ \cite{argo1}.

The two MAGIC IACTs, spaced away at a distance of 85 m, are among the
biggest instruments of this kind in operation and are located on La Palma
(Canary Islands). The MAGIC threshold for $\gamma$-rays is 50 GeV (but it
goes down to 25 GeV when the so-called SUM trigger system is employed) and
the energy resolution is $\sim$ 25\% above 200 GeV. In recent years MAGIC
gave several contributions to galactic and extragalactic astrophysics. One
example is the observation of the emission of the SuperNova Remnant
IC-443, at a distance of 1.5 Kpc (see fig. 1 of ref. \cite{magic2}): it
appears not coincident with the center of the SNR shell, but is correlated
with a maser emission. A possible source of this VHE radiation is
$\pi^\circ$ decay coming from CR accelerated in the dense molecular cloud.
The second example (see fig. 2 of ref. \cite{magic2}) is the measure of
the $\gamma$-ray flux of the X-ray binary system LSI +61 303, which
clearly shows the periodic nature of the emission. Very interesting is the
regular monitoring campaign of the giant elliptical radio galaxy M87, made
together with HESS \cite{hess} and VERITAS \cite{veritas}. On February 1,
2008, MAGIC detected a flare which reached a maximum of 15\% of the Crab
Nebula flux (see fig. 5 of ref. \cite{magic2}).

ARGO-YBJ \cite{argo1} is a collaboration between Italy and China located
in Tibet. The detector consists of modules of 12 RPCs, to form an inner
area of 5600 m$^2$, surrounded by 23 additional clusters (``guard ring").
The array works in two independent data acquisition modes: ``scaler" and
``shower" mode. In scaler mode, each module counts the rates of events
with a total number of hits $\geq 1$, $\geq 2$, $\geq 3$, $\geq 4$ every
0.5 s, without any direction information (GRB searches). In shower mode, a
valid event requires at least 20 particles registered within 420 ns, with
an angular resolution of 0.2$^\circ$ for a primary above 10 TeV and
2.5$^\circ$ at $\sim$ 100 GeV. Figure 1 of ref. \cite{argo2} shows the sky
map for events with $N_{PAD}\geq 40$, corresponding to 424 days of data
taking, after correcting for excesses in CR flux from the galactic
anticentre. The Crab Nebula and Mrk 421 are detected with statistical
significance 7.0 and 8.0, respectively. In fig. 5 (left) of ref.
\cite{argo3} it is reported the distribution of statistical significance
of 26 GRB detected by satellites in the period July 2006 - July 2007 and
November 2007 - January 2009, showing no excess neither as prompt nor
prior/delayed emission. An important contribution of this experiment was
also the measure of the production cross-section of protons and air nuclei
\cite{argo4}, which follows from the angular distribution of showers.

\section{High energy experiments}

Gamma ray astronomy at high energies has to face the problem of the photon
pair production over the infrared and microwave backgrounds, which
restricts the distances over which gamma rays can travel without
attenuation. Neutrinos would be the best messenger particles, if not for
their very small cross-section with matter, which is an obstacle for their
efficient detection. High energy experiments usually rely on charged CR
detection. The biggest challenge of CR observations at high energy lies in
the fact that the flux of particles is very small, implying the need of
enormous exposures. Two detector techniques are used in this range of
energy: surface detector arrays on the ground (as Haverah Park or AGASA
\cite{arrays}), or air fluorescence detectors (as HiRes \cite{hires1}),
which collect the fluorescence light emitted by nitrogen molecules hit by
the secondary particles of an atmospheric shower. While in the first case
the experiment samples only the lateral distribution of particles at a
given atmospheric depth, relying on simulations for the determination of
mass and energy of the primary particle, in the second case one can infer
the longitudinal evolution of the shower in atmosphere, using the known
proportionality between air fluorescence and charged particle energy loss.
In particular, stereo configurations of fluorescence telescopes can be
used, allowing an angular resolution of less than 1$^\circ$ \cite{hires1}.
However, while a surface array operates continuously, a drawback of the
fluorescence technique is the low duty cycle of $\sim$ 10\%, due to the
fact that one needs dark, moonless nights with good atmospheric
conditions. The two types of detection techniques are complementary and
the new generation of experiments (Pierre Auger Observatory \cite{auger1},
TA\&Tale \cite{ta_tale}) uses both of them to check systematic errors.

At $6\cdot 10^{19}$ eV pion photoproduction of protons over the microwave
background radiation gives a suppression of the CR flux (nuclei
photodisintegration and gamma ray pair production are the corresponding
processes for the other CR components): this is the so-called
Greisen-Zatsepin-Kuz'min (GZK) effect \cite{gzk}. Other physics highlights
for UHE CRs are: their mass composition, the identification of their
sources, the presence of photons or neutrinos together with charged CRs.
Recent observations coming from new experiments allowed us to shed light
on these topics. In the following, for the sake of brevity, we will only
report some results of the Pierre Auger Observatory (PAO) experiment.

18 countries cooperate in the PAO experiment \cite{auger1} for building
two CR observatories, one in the Southern hemisphere, at Malarg\"{u}e in
Argentina, completed in June 2008, and one in the Northern hemisphere, in
Colorado, under development. Auger South consists of a Surface Detector
(SD) array of 1600 water Cherenkov, covering an area of about 3000 km$^2$
on a triangular grid with 1.5 km spacing, and a Fluorescence Detector (FD)
made by 24 optical telescopes in 4 buildings at four peripheral sites.

Figure 5 of ref. \cite{auger2} shows the energy spectrum of UHE CRs
derived from the combination, with a maximum likelihood method, of two
kind of data: hybrid data, that is events detected with both the FD and
the SD, which have a more accurate reconstructed energy (data were
collected between November 2005 and May 2008) and SD data (until December
2008). While the two sets have the same systematic uncertainty in the
energy, the normalization uncertainties are additional constraints in the
combination. The comparison with HiRes data shows some discrepancy, which
can be partly reconciled by an energy shift of the energy scale of the two
experiments. Results single out the break in the power law of the CR
spectrum called ``ankle" at log$_{10}$ ($E$/eV) = 18.61$\pm$0.01 and give
indication of the GZK suppression at log$_{10}$ ($E$/eV) = 19.61$\pm$0.03
with more than 20$\sigma$ statistical significance. The mass composition
of CR at different energies can be inferred by the so-called ``elongation
rate", that is the change of the average depth of the shower maximum,
$\langle X_{max}\rangle$, as a function of the energy. This is shown in
fig. 3 of ref. \cite{auger3} (left plot), together with its
shower-to-shower fluctuation, rms($X_{max}$) (right plot), a measure which
is possible only thanks to the excellent resolution of FD ($\sim$ 20
g/cm$^2$). The behaviors of $\langle X_{max}\rangle (E)$ and
rms($X_{max}$) give indication of an increasing average mass of the
primary particles with energy, somehow in contradiction with corresponding
data from HiRes \cite{hires2}. Some explanation was proposed for solving
the puzzle, for example the unexpected changes of the depth of first
interaction due to a rapid increase of cross section and/or increase of
inelasticity above $2\cdot 10^{18}$ eV, which enhance the role of biases
due to small statistics \cite{ww}. Finally, we briefly mention the PAO
results on the arrival directions of CRs, referring for details the
interested reader to the published papers \cite{auger4,auger5}. Table I of
ref. \cite{auger5} summarizes the results of the scan on the high energy
events ($E>55$ EeV) detected by the Auger experiment in the period 1
January, 2004 through 26 May, 2006 (exploratory period or Period I) and 27
May, 2006 through 31 March, 2009 (Periods II and III). A correlation over
angular scales of less than 6$^\circ$ with directions towards nearby AGNs,
listed in the V\'eron-Cetty and V\'eron catalog, is established, even if
its degree seems to be weaker than suggested by earliest data (Period II).
While a clear interpretation of this signal has to wait for more data,
remarkable features are that the values of the parameters that
characterize the correlation are stable with time and, in particular, the
threshold energy which maximizes it coincides with the GZK suppression
one.

\section{Future prospects}


IACT technique is very promising for future experiments, since it has the
potential for large improvements, like the increase of flux sensitivity,
at the level of $\sim 10^{-3}$ the Crab Nebula flux, by increasing the
collection area from 0.1 km$^2$ to $\geq$ 1 km$^2$ with the concept of
large arrays of IACTs. This enhancement, together with its good angular
resolution, will allow joint studies with satellite experiments, like
Fermi-LAT \cite{fermi}, which will help, for example, in breaking the
degeneracy of leptonic and hadronic emission models. However, $\gamma$-ray
astronomy requires also new advanced EAS array experiments, like HAWC
\cite{hawc}, since these can view continuously a large region of the sky.

At higher energies, some enhancements are proposed for existing
experiments, like areas of denser detector arrays, called ``infill", for
surface detectors, or extra fluorescence telescopes for enlarging the
field of view of fluorescence detectors, both at the aim of going to lower
energies, or/and the use of radio-detection technique. These will be very
helpful in answering to some of the unknown questions about CRs. In fact,
we can be confident in some physical results indicated by experiments,
like the existence of the GZK suppression at energies $>6\cdot 10^{19}$ eV
and the correlation between event directions and the super galactic plane.
However, several questions remain still uncertain, like the mass
composition at high energy or the identity of objects with which CR events
correlate, which we trust will be addressed by the near future UHE
cosmic-ray experiments.

\end{document}